\documentclass[prb,twocolumn,superscriptaddress,showpacs,amsmath,amssymb]{revtex4-1}

\usepackage{graphicx}
\usepackage{latexsym}
\usepackage{amsmath}
\usepackage{amssymb}
\usepackage{amsfonts}
\usepackage{color}
\usepackage{bm}
\usepackage{verbatim}

\definecolor{MS-color}{RGB}{128,0,128}

\usepackage{framed}
\definecolor{shadecolor}{RGB}{222,222,221}
\begin{document}


\title{Spin torques and magnetic texture dynamics driven by the supercurrent in superconductor/ferromagnet structures.}

 \date{\today}

\author{I. V. Bobkova}
\affiliation{Institute of Solid State Physics, Chernogolovka, Moscow
  reg., 142432 Russia}
\affiliation{Moscow Institute of Physics and Technology, Dolgoprudny, 141700 Russia}

\author{A. M. Bobkov}
\affiliation{Institute of Solid State Physics, Chernogolovka, Moscow reg., 142432 Russia}

 \author{M.A.~Silaev}
 \affiliation{Department of
Physics and Nanoscience Center, University of Jyv\"askyl\"a, P.O.
Box 35 (YFL), FI-40014 University of Jyv\"askyl\"a, Finland}


 \begin{abstract}
 We introduce the general formalism to describe spin torques induced by the
 supercurrents injected from the adjacent superconducting electrodes into the
 spin-textured ferromagnets.
  By considering the adiabatic limit for the equal-spin superconducting correlations in the ferromagnet
 we show that the supercurrent can generate both the field-like spin transfer torque and the spin-orbital
 torque. These dissipationless spin torques are expressed through the  current-induced corrections to the effective field
 derived from the system energy.
 The general formalism is applied to show that the supercurrent can either shift or move the magnetic domain walls
 depending on their structure  and the type of spin-orbital interaction in the system.
 These results can be used for the prediction and interpretation of the experiments studying
 magnetic texture dynamics in superconductor/ferromagnet/superconductor Josephson junctions and other hybrid structures.
   \end{abstract}

  \maketitle

 \section{Introduction}

  It has been commonly recognized that reducing Joule heating effects and power consumption
  are among the main priorities for the development of  electrically controlled magnetic memory devices
  \cite{Brataas2012, Locatelli2013,ChenJune,Yamaguchi2005}.
  Since the first spin transfer torque (STT) experiments  \cite{Katine2000,Myers1999}
  much effort has been invested to optimize the switching currents,  thermal stability and tunnel
  magnetoresistance of the magnetic tunnel junctions \cite{Huai2004,ChenJune,Wang2011}.
  Thermal effects are also of the crucial importance for the operation of the
   other type of STT memory -  magnetic racetrack memory
  \cite{Parkin2008,Hayashi2008,Allwood2005} based on the electrical control over the domain wall (DW) motion.
  The progress in improving  these spin memory devises depends crucially
  on the competition between the thermal stability of DWs and large  current densities
  required to overcome the pinning forces \cite{Vernier2004,Yamaguchi2004,Yamaguchi2005,Zhang2004,Tatara2004,Thiaville2005,Obata2008,Miron2011}.
 As an alternative route to the low-power manipulation
 of magnetic textures the current-driven magnetic skyrmion dynamics has attracted large interest  \cite{Nagaosa2013,Iwasaki2013,Fert2013}.

In applications which require  very large currents, for example in  powerful magnets, using superconducting materials
have been proven to be an effective solution to eliminate Joule heating effects.
In view of the energy-saving spintronics it is quite appealing to
employ the spin torques generated by the dissipationless spin-polarized superconducting currents (supercurrents).
 The existence of spin-polarized supercurrents is ubiquitous to the spin-textured
 superconductor/ferromagnet (SC/FM) hybrid structures resulting from
 long-range spin-triplet proximity \cite{Bergeret2001, Bergeret2004, Bergeret2005b, Bergeret2005}.

 Recently there have been many works studying spin-polarized supercurrents
 in various SC/FM systems (for the review see Refs.\citep{Eschrig2015a, Linder2015}).
However the supercurrent-induced spin torques have been characterised theoretically only in several model systems:
in Josephson junctions through the single-domain magnets \cite{Zhu2004,Nussinov2005,Holmqvist2011, Buzdin2008, Konschelle2009},
two\cite{Waintal2002,Linder2011,Halterman2016} and three\cite{Kulagina2014} FM layers, in ferromagnetic spin-singlet\citep{Linder2012}
and spin-triplet superconductors \cite{Takashima2017}.
The general understanding of the supercurrent- spin texture interaction has been lacking
since there is no direct connection between the above examples and practically interesting systems - bulk
non-homogeneous FMs. That is, the possibility of moving DWs and skyrmions by injecting the
supercurrent in real ferromagnets has been an open question for a long time
despite of the large attention to the subject.

This challenging question is addressed in the present paper.
We employ the adiabatic approximation which is widely used for the description of kinetic processes in
metallic ferromagnets with spin textures
including the calculation of conductivity \cite{Nagaosa2013}
and spin transfer torques\cite{Berger1996} in the inhomogeneous FMs.
We bring this approach to the realm of superconducting systems to describe their
transport properties governed by equal-spin superconducting correlations.
For that we go beyond the commonly used quasiclassical theory
of hybrids\cite{Buzdin2005, Bergeret2005} which has been designed to treat only the weak ferromagnets with the exchange
splitting much less than the Fermi energy.
Instead of that we employ the recently developed approach of generalized quasiclassical theory\citep{Bobkova2017a} which
allows for the description of proximity effect in strong ferromagnets with the exchange splitting much larger than other
energy scales and comparable to the Fermi energy.

We show that the spin-polarized superconducting current can induce
magnetization dynamics, described in general by the Landau-Lifshitz-Gilbert (LLG) equation
 %
 %
  \begin{equation} \label{Eq:LLG0}
 \bm{\dot M} = - \gamma \bm M\times \bm H_{eff} + \frac{\alpha}{M} \bm M \times \bm{\dot M} ,
 \end{equation}
where $\gamma = 2\mu_B$ is the electron gyromagnetic ratio. The second term in the r.h.s. is the Gilbert damping.
The superconducting spin current $\bm J$ can induce two types of spin torques which can be written as the
correction to effective field $-\gamma \bm M \times \tilde{\bm H}_{eff} = \bm N_{st} + \bm N_{so}$.
The first term here is the
adiabatic spin-transfer torque \cite{Slonczewski1996,Tatara2004,Koyama2011}while the second term $\bm N_{so}$ is the spin-orbital (SO) torque\cite{Miron2010,Gambardella2011}.
The non-adiabatic (antidamping) STT\cite{Zhang2004} is not produced by the supercurrent since it
breaks the time-reversal
symmetry of LLG equation and
can be considered as a correction to the dissipative Gilbert damping\cite{Garate2009}.
That is the antidamping STT should be connected with the quasiparticle contribution  which is beyond
the scope of our present study.


The paper is organized as follows.
In Sec.\ref{Sec:SpinToruesGeneral} the general equations for spin dynamics and spin torques generated by supercurrent are considered.
In Sec.\ref{Sec:DerivationST} we derive expressions for the spin-transfer and spin-orbital torques using the generalized quasiclassical theory.
In Sec.\ref{Sec:DerivationJE} we derive Josephson energy in SFS junctions and use it to provide an alternative derivation
of supercurrent spin torques. Sec.~\ref{Sec:textureDynamics} is devoted to the DW dynamics in SFS Josephson junctions induced by the supercurrent spin torques.
Our conclusions are given in Sec.\ref{Sec:Conclusions}.

\section{Spin torque generated by the supercurrent}
\label{Sec:SpinToruesGeneral}

 We use s-d model with the localized magnetization $\bm M$ and that of the itinerant electrons
 $\bm M_s=-\mu_B \bm s$, where $\bm s$ is the electron spin and $\mu_B$ is the Bohr magneton.
 The dynamics of localized spins is determined by
 the usual LLG equation with the contribution to the effective field resulting from exchange interaction
 with conductivity electrons
 \cite{TsymbalZuticBook}
 \begin{equation} \label{Eq:Dspin}
 \bm{\dot M} = - \gamma \bm M\times \bm H_{eff} + \frac{\alpha}{M} \bm M \times \bm{\dot M}  -  J_{sd} \bm M\times \bm M_s .
 \end{equation}
 The last term here is the source of spin torque and should be
 found from the kinetic equation for conductivity electrons.

 The  kinetic theory for the conduction electrons in metals can be formulated in terms of the
 matrix Green's function  $\check G= \check G(\bm r_1, \bm r_2, t_1,t_2)$ which has the following
 explicit structure in the Keldysh space
 $
  \check G =
  \left(
  \begin{array}{cc}
  \check G^R & \check G^K \\
  0 & \check G^A \\
  \end{array}
  \right),
 $
  %
  where $\check G^{R/A/K}$ are the retarder/advanced/Keldysh components.
  The general quantum kinetic equation reads:
  \begin{align} \label{Eq:Gorkov1}
  & i \{ \partial_t, \check G \}_t - [\hat H , \check G ]_{t,\bm r}  = \check I
  \\ \label{Eq:G0}
  & \hat H (t,\bm r)= -\frac{ \hat{\bm\Pi}_{\bm r}^2}{2m_F} +
  \left( \bm{ \hat\sigma} {\bm h} (\bm r,t) \right)\hat\tau_3 - i (\bm{\hat \sigma} \hat B \hat{\bm\Pi}_{\bm r} ) ,
  \\  \label{Eq:CollisionIntegral}
  & \check I = (\check \Sigma \circ \check G -  \check G\circ \check \Sigma) (\bm r_1,\bm r_2,t_1,t_2) .
  \end{align}
  Here we define the $\circ$-product as
 $(\hat A \circ \hat B) (t_1,t_2)= \int_{-\infty}^{\infty} dt \hat A(t_1,t)\hat B(t,t_2)$. The commutator is defined as $[\hat H, \check{G}]_t = \hat H(t_1,\bm r_1)\check{G} -\check{G} \hat H(t_2,\bm r_2)$,
  $\hat{\bm\Pi}_{\bm r} = \nabla - ie \hat \tau_3\bm A (\bm r) $,
  $\hat \sigma_i$ and  $\hat \tau_i$ are Pauli matrices in spin and Nambu spaces, respectively.
  The exchange field is determined by localized moments
  $\bm h =  - J_{sd} \bm M /2\mu_B$.
 The last term in Eq.(\ref{Eq:G0}) is the general form of a linear in momentum spin-orbit coupling (SOC) determined by
 the constant tensor coefficient $\hat B$.
 The collision integral in the r.h.s. of Eq.(\ref{Eq:Gorkov1}) is given by the Eq.(\ref{Eq:CollisionIntegral}).
  The self-energy term $\check \Sigma$ includes the effects related to disorder scattering as well as the
 off-diagonal superconducting self-energies.

  The conduction electron spin polarization $\bm s$, charge $\bm j$ and spin {$\bm J^i$} currents are given by
  \begin{align}
  & \bm s (\bm r,t) = -\frac{i}{8}{\rm Tr}_4 [ \bm{\hat \sigma} \hat\tau_3 \hat G^K]|_{\bm r_{1,2}=\bm r, t_{1,2}=t}.
  \label{m}
  \\ \label{j}
  & \bm j (\bm r, t) =
  {\rm Tr}_4 \left[   \frac{( \bm {\hat \Pi}_{\bm r_1} - \bm {\hat \Pi}_{\bm r_2}) }{8m_F}
  \hat\tau_3 \check G^K \right]|_{\bm r_{1,2}=\bm r, t_{1,2}=t}
  \\
  & \bm J^k (\bm r, t) =
  {\rm Tr}_4 \left[ \frac{( \bm {\hat \Pi}_{\bm r_1} - \bm {\hat \Pi}_{\bm r_2}) }{16m_F}
  \hat\sigma_k\check G^K \right]|_{\bm r_{1,2}=\bm r, t_{1,2}=t} .
  \label{Js}
  \end{align}

 The strategy of studying magnetization dynamics consists of solving the coupled LLG (\ref{Eq:Dspin}) and kinetic equations
 (\ref{Eq:Gorkov1})-(\ref{Eq:CollisionIntegral}) together with the expression for the
 magnetic moment (\ref{m}).
 However, the general problem is too complicated for the analysis.
 In the next section  (\ref{Sec:DerivationST}) we discuss the simplification of the kinetic equation using the so-called generalized
 quasiclassical approximation \cite{Bobkova2017a} adopted to treat the non-stationary problems.

 Besides that, significant simplification can be obtained in the linear response limit when the dynamics
 of magnetization is slow so that the characteristic frequency is small as compared to the energy gap in the quasiparticle spectrum.
 In this case we can make use of the quasi-stationary  equation for the electron magnetization which
 is obtained from Eq.(\ref{Eq:Gorkov1}).
 Multiplying it by $\bm{\hat \sigma} $ from the left and  taking the trace we obtain
  \begin{align} \label{m_equation}
  &  -\partial_t \bm s  + \nabla_j \bm J_{j} =  \\ \nonumber
  & \frac{J_{sd}}{\mu_B}  (\bm M \times \bm M_s) + 2 m_F (\bm B_j \times \bm J_{j})
  + \frac{{\rm Tr}_4 [\hat{\bm \sigma} \check{I}]^K }{8} .
  \end{align}
  Here we introduce the vector $\bm B_j = (B_{xj},B_{yj},B_{zj})$, which is determined by $j$-th
  { coordinate} component of the
  tensor $\hat B$ and { $\bm J_j = (J_{j}^x, J_{j}^y, J_{j}^z)$ is the}  
  $j$-coordinate component of the spin current (\ref{Js}).
   Next, the driving term in the LLG equation (\ref{Eq:Dspin}) can be found neglecting
   the term with time derivative  in Eq.(\ref{m_equation})
   \begin{equation} \label{torque}
  \frac{J_{sd}}{\mu_B} (\bm M \times \bm M_s)
   =  \nabla_j \bm J_{j} -   2 m_F (\bm B_j \times \bm J_{j})
  - \frac{{\rm Tr}_4 [\hat{ \bm \sigma}\check{I}]^K }{8}.
  \end{equation}

  In this work we are interested in the quasi-equilibrium spin torques generated solely by the supercurrent without the
  contribution of non-equilibrium quasiparticles. That means the normal component of the current and
  the electric field are assumed to be absent.
   Generally, the last term with the collision integral in Eq.(\ref{torque}) has contributions
   both from the off-diagonal order parameter and the spin-orbital scattering self-energy.
   The present work is based on the following simplifying assumptions allowing to put
   ${\rm Tr}_4 [\hat{ \bm\sigma} \check{I}]^K =0$.
   First,
   we are interested in the spin torques occurring in the normal metal ferromagnetic interlayer
   where the order parameter is absent. Second, the exchange splitting between spin subbands is assumed to be
   large enough to suppress spin-flip transitions between them.
 Below we demonstrate that in this regime the first term in the r.h.s. of Eq.~(\ref{torque})
 produces the adiabatic  STT\cite{Slonczewski1996,Berger1996,Ralph2008},
 while the second term yields the spin-orbit torque
 \cite{Manchon2008,Miron2010,Gambardella2011}.
 In order to find these contributions we  calculate the spin supercurrent through the spatially-inhomogeneous ferromagnet.
 In the next section it is shown that in the adiabatic limit valid for the description of
 strong ferromagnet this calculation can be done analytically in the in the most general way  using the
 technique developed in Ref.\cite{Bobkova2017a}.

  \section{Generalized quasiclassical theory}
  \label{Sec:DerivationST}

  {\it Eilenberger equation for equal-spin correlations.}
   To find the results in the adiabatic approximation it is convenient to work in the local reference frame,
  where the spin quantization axis is aligned with the local direction of the exchange field in the ferromagnet.
  Then we use the transformation
  $\check{G}_{loc} = \hat U^\dagger \check G \hat U$ where
  $\hat U= \hat U ({\bm r,t}) $ is in general the time- and space-dependent unitary $2\times 2$ matrix that
   rotates the spin quantization axis $\bm z$ to the local frame determined by the exchange field, so that
    ${\bm h} \parallel {\bm z}$.
  To implement the adiabatic approximation we introduce the
 equal-spin pairing components of the GF
  \begin{equation} \label{Eq:GFProjection}
 \hat G^\sigma_{ES} = \frac{1}{4}\sum_{i} \hat\tau_i {\rm Tr} [\check\gamma_{\sigma i}\check G_{loc}] .
 \end{equation}
 Here the projection operators to spin-up and spin-down states defined by the index $\sigma=\pm 1$
 are given by
 $\check\gamma_{\sigma 0} = \hat\tau_0 \hat\sigma_0 + \sigma \hat\tau_3 \hat\sigma_3 $,
 $\check\gamma_{\sigma1} = \hat\tau_1 \hat\sigma_1 - \sigma \hat\tau_2 \hat\sigma_2 $,
 $\check\gamma_{\sigma 2} = \hat\tau_2 \hat\sigma_1 + \sigma\hat\tau_1 \hat\sigma_2 $,
 $\check\gamma_{\sigma 3} =  \hat\tau_3 \hat\sigma_0 + \sigma \hat\tau_0 \hat\sigma_3$.
 %
  %
  The generalized quasiclassical theory is formulated in terms of the spin-less propagators
  \begin{equation}\label{Eq:DefinitionQuasiclassics}
  \hat g_{\sigma} (\bm n_p, \bm r) =  -  \oint \frac{d\xi_{p\sigma}}{\pi i} \hat G^{\sigma}_{ES}(\bm p,\bm r) ,
  \end{equation}
  where $\hat G^{\sigma}_{ES} = \hat G^{\sigma}_{ES}(\bm p,\bm r)$ is the
  GF in the mixed representation, $\xi_{p\sigma} = p^2/2m_F + \sigma h - \mu$ and the notation
  $\oint$ means that the integration takes into account the poles of GF near the corresponding Fermi surface.
  Then in the adiabatic approximation, which neglects the
  coupling between equal-spin and mixed-spin correlations\cite{Bobkova2017a} we obtain the  generalized Keldysh-Eilenberger equation
  \begin{align}
  \label{Eq:Eilenberger}
  & i\{\hat \tau_3\partial_t, \check{g}\}_t  +
  i {\bm v}_{\sigma}  \hat\partial_{\bm r} \hat g_{\sigma} -
  [ \hat\Sigma_{\sigma}, \hat g_{\sigma} ]_t =0 ,
  \\
  \label{Eq:GradientCovariant}
  & \hat \partial_{\bm r} = \nabla - ie[\bm A\hat \tau_3, .]_t + i\sigma [\bm Z\hat \tau_3, .]_t .
  \end{align}
  Here the spin-dependent Fermi velocities $v_\pm = \sqrt{2(\mu \pm h)/m_F}$ are
 determined on each of the spin-split Fermi surfaces .
 The spin-dependent gauge field is given by the superposition of two terms
 $\bm Z= \bm Z^{m} + \bm Z^{so}$, where $Z^{m}_i = -i {\rm Tr} \Bigl( \hat \sigma_z \hat U^\dagger \partial_i \hat U \Bigr)/2 $ is the texture-induced part and
 the term $Z^{so}_i = m_F (\bm{m }\bm B_i)$ (where $\bm m = \bm M/M$) appears due to the SOC.

 One can see that the Eilenberger-type equations for the spin-up/down correlations
 contain an additional U(1) gauge field ${\bm Z}$ which is added to the usual electromagnetic vector potential ${\bm A}$
 with the  opposite effective charges for spin-up and spin-down Cooper pairs.
 On a qualitative level it is equivalent to the adiabatic
 approximation in the single-particle problems that allows to describe the quantum system evolution in terms of the Berry
 gauge fields\cite{Bliokh2005}.

 {\it Charge and spin currents.}
  The Eilenberger equations (\ref{Eq:Eilenberger}) are supplemented by the
   expressions for the charge
   current $\bm j$ and the spin current $\bm J^k$, where $k$ denoted the spin index.
  The former is given by
  \begin{equation} \label{Eq:ChargeCurrent}
  {\bm j} (t)= -\frac{\pi e}{4}   \sum_{\sigma=\pm}  \nu_{\sigma}
  \langle {\bm v}_{\sigma} {\rm Tr} [ \hat\tau_3\hat g_{\sigma}^K (t,t)] \rangle ,
  \end{equation}
  where $\nu_{\sigma}$ are the spin-resolved DOS and $ \langle .. \rangle$
  denotes the averaging over the spin-split Fermi surface.
  The spin current in rotated frame is given by
  \begin{equation} \label{Eq:SpinCurrent}
  \bm{\tilde J}^z(t) = -\frac{\pi}{8} \sum_{\sigma=\pm} \sigma  \nu_{\sigma}
  \langle {\bm v}_{\sigma} {\rm Tr} [ \hat\tau_3\hat g_{\sigma}^K (t,t)] \rangle
  \end{equation}


{\it Diffusive limit.}
 %
 Let us consider the system with large nonmagnetic impurity scattering rate as compared to the
 superconducting energies determined by the bulk energy gap $\Delta$.
 In this experimentally relevant diffusive limit it is possible to derive the generalized
 Usadel theory with the help of the
 normalization condition $(\hat g_\sigma \circ \hat g_\sigma) (t_1,t_2)= \hat \delta (t_1 - t_2)$ which holds due to the commutator structure of the
 quasiclassical equations (\ref{Eq:Eilenberger}).

 The impurity self-energy in the Born approximation is given by
 $\hat\Sigma_{\sigma} =
 \langle \hat g_{\sigma} \rangle/2i\tau_{\sigma}$.
 In the dirty limit we have
 \begin{equation} \label{Eq:Usadel}
 2\tau_{\sigma}
 ({\bm v}_{\sigma} \hat\partial_{\bm r}) \hat g_{\sigma} =
 - [ \langle \hat g_{\sigma} \rangle , \hat g_{\sigma} ]_t.
 \end{equation}
The solution of Eq.~(\ref{Eq:Usadel}) can be found as
$\hat g_{\sigma}= \langle \hat {g}_{\sigma} \rangle +\hat {\bm g}_{\sigma}^a \bm p_\sigma/p_\sigma$,
where the anisotropic part of the solution $\hat {\bm g}_{\sigma}^a$ is small with respect to $ \langle \hat {g}_{\sigma} \rangle$.
Making use of the relation $\{ \langle \hat g_{\sigma} \rangle , \hat {\bm g}_{\sigma}^a \}_t =0$,
which follows from the normalization condition, one obtains
 \begin{equation}\label{Eq:UsadelExpansion}
 \hat {\bm g}_{\sigma}^a =
 - \tau_{\sigma}
 {\bm v}_{\sigma} \langle \hat g_{\sigma} \rangle \circ
 \hat\partial_{\bm r} \langle \hat g_{\sigma} \rangle.
 \end{equation}

 Substituting to Eq.(\ref{Eq:Eilenberger}) and omitting the angle brackets we get the diffusion equation
 \begin{equation} \label{Eq:UsadelEquation}
 \{\hat \tau_3\partial_t, \hat{g}_\sigma\}_t  -
 D_{\sigma} \hat\partial_{\bm r}( \hat g_{\sigma}\circ \hat\partial_{\bm r} \hat g_{\sigma})  =0 ,
 \end{equation}
 where $D_{\sigma}$ are the spin-dependent diffusion coefficients, in the
 isotropic case given by  $D_{\sigma} = \tau_{\sigma}  v^2_{\sigma}/3$. This equation is a spin-scalar equation, but cannot describe conventional spin-singlet superconducting correlations unlike the standard spin-scalar form of the non-stationary Usadel equation \cite{LarkinOvchinnikov}. It is only applicable for strong ferromagnets and describes equal-spin triplet correlations residing at one and the same Fermi-surface. Therefore, this equation is a non-stationary generalization of the corresponding equations for homogeneous strong ferromagnets \cite{Grein2009} and inhomogeneous strong ferromagnets \cite{Bobkova2017a}.

 The current and spin current are obtained substituting expansion (\ref{Eq:UsadelExpansion}) to
 Eqs.(\ref{Eq:ChargeCurrent})-(\ref{Eq:SpinCurrent})
 \begin{align} \label{Eq:ChargeCurrentDiff}
  & {\bm j} =   \frac{\pi e}{4} \sum_{\sigma =\pm}
  \nu_\sigma D_{\sigma} {\rm Tr}[ \hat\tau_3 \hat g_{\sigma} \circ \hat\partial_{\bm r}\hat g_{\sigma} ]
  \\  \label{Eq:SpinCurrentDiff}
  & \bm{\tilde J}^z = \frac{\pi}{8} \sum_{\sigma =\pm  } \sigma
  \nu_\sigma D_{\sigma} {\rm Tr}[ \hat\tau_3 \hat g_{\sigma}\circ \hat\partial_{\bm r}\hat g_{\sigma} ]
  \end{align}
  Further simplification can be obtained as follows.
  First, due to the normalization condition
    we introduce the parametrization of Keldysh component in terms of the distribution function
  $\hat g^K_\sigma = \hat g^R_\sigma \circ \hat f_\sigma- \hat f_\sigma\circ \hat g^A_\sigma$.
  Then, switching to the mixed representation in time-energy domain
  $\hat{g}_\sigma (t_1,t_2) = \int_{-\infty}^{\infty} \hat{g}_\sigma(\varepsilon, t)
  e^{-i\varepsilon (t_1-t_2) } d\varepsilon/2\pi$, where $t=(t_1+t_2)/2$ we keep only the
  lowest order terms in the time derivatives.

  As an example of the above procedure one can obtain from (\ref{Eq:ChargeCurrentDiff}) the charge current in the normal state
  $\bm j = e^2 (\nu_+D_+ - \nu_-D_-)\bm E_e$ driven by the
  emergent electric field\cite{Volovik1987,Nagaosa2013}  $\bm E_{e} = - \partial_t \bm Z$. We however will neglect these effects
  and take into account only the quasi-equilibrium contributions to the currents given by
  \begin{align} \label{Eq:ChargeCurrentDiff_QE}
  & {\bm j} = \sum_{\sigma =\pm} \frac{e\nu_\sigma D_{\sigma}}{8}
   \int_{-\infty}^{\infty} d \varepsilon f_0
  {\rm Tr} (\hat \tau_3 \hat{\bm J}^{RA}_\sigma )
  \\  \label{Eq:SpinCurrentDiff_QE}
   & \bm{\tilde J}^z = \sum_{\sigma =\pm} \frac{\sigma\nu_\sigma D_{\sigma}}{16}
   \int_{-\infty}^{\infty} d \varepsilon f_0
  {\rm Tr}( \hat\tau_3 \hat{\bm J}^{RA}_\sigma ),
  \end{align}
 where $\hat{\bm J}^{RA}_\sigma = \hat g^{R}_{\sigma} \hat\partial_{\bm r}\hat g^{R}_{\sigma} -
 \hat g^{A}_{\sigma} \hat\partial_{\bm r}\hat g^{A}_{\sigma} $ is the spectral current
 and $f_0(\varepsilon) = \tanh (\varepsilon/2T)$ is equilibrium distribution function.


 {\it Supercurrent-induced torque.}
 In the quasi-equilibrium regime when the time derivative of the GF in the mixed representation can be neglected
 Eqs.(\ref{Eq:Eilenberger}) or (\ref{Eq:UsadelEquation}) yield the conservation of spin current in rotating frame
 $\nabla\cdot\bm{\tilde J}_z = 0$.
 The spin current in the laboratory frame is  given by ${\bm J}^k =  R_{kz} \bm{\tilde J}^z$  which can be written in the form
  \begin{equation} \label{Eq:LabSpinCurrent}
  {\bm J}^k (\bm r) =  m_k (\bm r) \bm{\tilde J}^z .
  \end{equation}
  It is not conserved due to the spatially-dependent magnetization of d-electrons  $\bm m = \bm m(\bm r)$.

 Substituting Eq.~(\ref{Eq:LabSpinCurrent}) into Eq.(\ref{torque})
 we obtain the torque, induced by the supercurrent in the quasiequilibrium regime:
   \begin{align} \label{torque_adiabatic}
   & J_{sd} \bm M_s\times \bm M = \bm N_{st} + \bm N_{so},
   \\ \label{torque_st}
   & \bm N_{st} =  2 \mu_B (\bm{\tilde J}^z \nabla) \bm{ m},
   \\  \label{torque_so}
   & \bm N_{so} = 4 \mu_B m_ F( \bm{ m}\times \bm B_j) {\tilde J}^z_{j}.
  \end{align}
    Here
  $\bm N_{st}$
  is the supercurrent spin transfer torque, which takes only the form of the adiabatic torque in the considered approximation,
  and $\bm N_{so}$ is the spin-orbit torque. Its particular structure strongly depends on the type of
  the spin-orbit coupling, realized in the system.
  Below we show that due to the
  coherent nature of the spin-polarized superconducting current
    the same result can be obtained from the energy functional of the system yielding the correction to the effective field.

 \section{Supercurrent spin torques as corrections to the effective field}
 \label{Sec:DerivationJE}
 Above we have derived general expressions (\ref{torque_st},\ref{torque_so}) for the superconducting
 spin torques starting from the kinetic equation treated in the adiabatic limit.
 For any particular system one can find the spin torques solving generalized Eilenberger/Usadel equations for the
 quasiclassical propagators and calculating the spin current according to Eq.(\ref{Eq:SpinCurrentDiff}).

 An alternative approach to obtain superconducting spin torques is based on the description of
 magnetization dynamics in terms of the phenomenological expression for the effective  field
 $\bm H_{eff} = - \delta F/\delta \bm M$, where $F=F(\bm M)$ is
the system energy  as a functional of the magnetization distribution.
The LLG equation without dissipation terms is given by
 \begin{equation} \label{Eq:LLG}
  \bm{ \dot M }= -\gamma \bm M\times \bm H_{eff} .
 \end{equation}
  This approach cannot be applied to derive spin transfer torques in the normal state where
 the conduction electron magnetization is not coherent.
 In contrast to the normal system superconducting electrons are in the macroscopically coherent state.
 Therefore the total energy of the system written in terms of the macroscopic variables
 describes the interaction between the condensate spin and
 the ferromagnetic order parameter.

 Based on the above discussion one can conclude that the superconducting spin transfer torques (\ref{torque_st}) and (\ref{torque_so})
 can be obtained from the energy arguments.
 To demonstrate this we consider a generic example of the
  Josephson system consisting of superconducting leads coupled through the ferromagnet
  with non-homogeneous magnetization texture.
  In general this task is rather complicated and requires extensive numerical calculations
  for each particular system considered.
  However, in strong ferromagnets the general expressions for Josephson spin and charge currents
  for different magnetic textures of the interlayer can be obtained
  using the machinery of the generalized quasiclassical theory \cite{Bobkova2017a}.


 We consider the 1D magnetic texture $\bm M= \bm M(x)$ in the interlayer of the thickness $d$ between two
 superconducting interfaces, located at $x= \pm d/2$. The superconducting order parameter phase difference
 between them is $\chi$.  The current-phase relation for this setup has been found\cite{Bobkova2017a} as the superposition
 of partial currents carried by the spin-up and spin-down Cooper pairs:
 \begin{equation} \label{josephson}
 j (\chi) = \sum_{\sigma =\pm} j_\sigma \sin\left(\chi+2 \sigma  \int_{-d/2}^{d/2} Z_x dx \right).
 \end{equation}
 The amplitudes $j_\sigma$ are determined by the boundary conditions at FM/SC interfaces
 and the overlap factor of the equal-spin correlations injected from the  opposite SC electrodes
 $j_{\sigma} \propto e^{-d/\xi_{N\sigma}}$,  where $\xi_{N\sigma} = \sqrt{D_\sigma/T}$ is the
 spin-dependent normal metal correlation length \cite{Bobkova2017a}. The other characteristic scale of the problem is the characteristic length of the magnetic inhomogeneity. In the case of the domain wall it is the wall size $d_w$. If we are interested in the domain wall motion and consider the situation when the DW is located inside the interlayer ($d_w < d$) and not in the vicinity of S/F interfaces, the amplitudes $j_\sigma$ do not depend on $d_w$ at all. But this scale enters the Josephson current via the effective gauge field $\bm Z = -m_x (m_y\nabla m_z - m_z\nabla m_y)/2m_\perp^2$, where $m_\perp =\sqrt{ m_y^2 + m_z^2}$. $\bm Z$ is a crucial factor giving rise to the DW dynamics, as it is explained below. In more general case, when the DW is wide $d_w>d$ or the DW is located in the vicinity of a S/F interface, the amplitudes $j_\sigma$ also depend on $d_w$ via boundary conditions \cite{Bobkova2017a}, but consideration of the DW dynamics presented below is not applicable in this case.

 The rotated-frame spin current $\tilde{J}^z \equiv \tilde J_{x}^z $ is given by the difference of the partial
 spin-up/down currents
 \begin{equation}
 \label{josephson:Spin}
 \tilde{J}^z (\chi) =
 \frac{1}{2e}\sum_{\sigma =\pm} \sigma j_\sigma \sin\left(\chi+2 \sigma \int_{-d/2}^{d/2} Z_x dx\right).
 \end{equation}
  The Josephson energy can be obtained according to the usual relation
  \begin{equation}\label{Eq:JosephsonEnergy}
  F_J =  const - \frac{1}{2e}\sum_{\sigma =\pm} j_\sigma
   \cos\left(\chi+2 \sigma  \int_{-d/2}^{d/2} Z_x dx \right).
   \end{equation}
  The current-phase relation (\ref{josephson}) is given by $j (\chi) =2e (dF_J/d\chi)$.
  Therefore calculating the correction to effective field $\tilde{\bm H}_{eff} = - \delta F_J/\delta \bm M $
  (see details in Appendix\ref{Apx:1}) we obtain
  \begin{equation} \label{Eq:EffectiveField}
  \gamma \bm M\times \tilde{\bm H}_{eff} =  2 \mu_B \tilde J^z ( 2m_F\bm B_x\times \bm m - \nabla_x \bm m)
  \end{equation}
  Substituting the result  (\ref{Eq:EffectiveField}) to the LLG equation we get the
  spin transfer torques identical to the Eqs.(\ref{torque_adiabatic},\ref{torque_st},\ref{torque_so}).

  The above energy consideration demonstrates that  the
  direct coupling between the magnetization and superconducting current exist even in the limit when
 the spontaneous charge current is absent.
  Indeed,  the spontaneous phase shift of the Josephson current-phase relation (\ref{josephson}) is given by
  $\tan \chi_0= \tan\varphi (j_+-j_-)/ (j_+ + j_-)$, where $\varphi=2\int_{-d/2}^{d/2} Z_x dx$.
  Therefore $\chi_0=0$  in the limit when the spin subbands are formally degenerate
   $j_+ = j_-$. At the same time the Josephson spin current
   (\ref{josephson:Spin}) and correspondingly the spin torque in the Eq.(\ref{Eq:EffectiveField})
   are non-zero.
  This result generalizes the previously suggested mechanism of
  the supercurrent-induced spin-orbital torque stemming from the $\chi_0=\chi_0(\bm m)$ dependence\citep{Buzdin2008,Konschelle2009}.
  In our case it is not only the phase shift, but in addition the overall critical current in Eq.(\ref{josephson})
  which depends on the magnetization $j_c= \sqrt{j_+^2 + j_-^2 + 2j_+j_-\cos(2\varphi)}$ through $\varphi=\varphi(\bm m)$.
  This provides the non-zero effective field even in the case of degenerate bands.

 \section{Supercurrent driven magnetic texture dynamics}
 \label{Sec:textureDynamics}


 \subsection{General case of the texture dynamics driven by the adiabatic STT}
 \label{Sec:HelixRotation}

 The first striking consequence of the dissipationless supercurrent spin torques is the possibility to realize the
 quasiequilibrium magnetic texture dynamics driven solely by the adiabatic STT generated by the superconducting current.
 In the absence of dissipation the LLG equation (\ref{Eq:LLG0}) have the solution in the form of the travelling
 wave $\bm m = \bm m(x-ut)$ with the constant velocity determined by the spin current $u = 2\mu_B\tilde{J}^z/M$.
 For the  periodical magnetic structure , e.g. magnetic helix
 that yields locally rotating magnetization with  the frequency defined by $\omega\sim  u/L$,
 where $L$ is the period. However, these time-dependent quasiequilibrium solutions do not correspond to the
 ground state. It  can be reached only in the presence of the Gilbert damping which transforms the magnetic texture in such a way to
 compensate the effective field generated by the spin-polarized supercurrent. Therefore, eventually the systems will stop at the
 stationary state when $\bm H_{eff}=0$.
 In the absence of dissipation the same quantity $u$ determines the characteristic velocity of the domain wall motion by the adiabatic STT in the system. In principle, current-driven motion of DWs in Josephson junctions with strong ferromagnets can be realized in different systems with high enough critical current densities. High critical currents through strong ferromagnets are typically carried by equal-spin triplet correlations, which decay on the length scale $\xi_{N\sigma}$ inside the ferromagnet \cite{Bergeret2005}. The Josephson current carried through strong ferromagnets by equal-spin triplet pairs was experimentally reported in different systems \cite{Khaire2010,Martinez2016,Singh2016,Witt2012,Robinson2010} (see also Ref.~\onlinecite{Eschrig2015a} for review), which are the promising elements for the dissipationless superconducting spintronics. Here we can estimate $u$ for the parameters of half-metallic CrO$_2$ nanostructures  \cite{Singh2016}.
 The maximal Josephson current density through the CrO$_2$ nanowire is $j_c\sim 10^9\; A/m^{2}$, which determines the spin current
 $\tilde{J}^z = j_c/(2e)$.  Taking into account the saturation magnetization $M = 4.75\times 10^5\; A/m$ we get the speed of the order of
 $u= 1\; m/s$.
 As we show below, in case if the initial state contains DW the
 ground state modified by the supercurrent can correspond either to the distorted DW or
 to the homogeneous state when the DW is eliminated from the sample. The dynamics of the initial state containing a DW under the applied supercurrent in the presence of the Gilbert damping in the LLG equation is also considered below.

 \subsection{Domain wall motion }
 \label{Sec:DWmotion}

 Now we consider the magnetic texture of the ferromagnet in the form of the DW.
 We are interested in its dynamics induced by the supercurrent spin torques, discussed above.
 The two particular types of DW are considered: head-to head DW and Neel DW.

 The particular shape of the DW is dictated by the combination of the anisotropy energy and the exchange energy.
 We start with the head-to-head DW. In this case
 the corresponding energy term can be written as follows:
 \begin{eqnarray}
 F = \frac{1}{2}\int d^3 r \left[ K_\perp m_y^2 -K m_x^2 +  A_{ex} ( \nabla_x \bm m )^2 \right],
 \label{anisotropy}
 \end{eqnarray}
 where $K>0$ and $K_\perp>0$ are the anisotropy constants for the easy and hard axes, respectively.
 $A_{ex}$ is the constant describing the inhomogeneous part of the exchange energy.
 The effective magnetic field
  $\bm H_{eff} = (1/M)(K m_x \bm x - K_\perp m_y \bm y +  A_{ex} \nabla^2_x \bm m)$.
 %
 It is convenient to parametrize the magnetization as follows:
 \begin{equation}
 \bm  m=  (\cos \theta, \sin \theta \cos \delta, \sin \theta \sin \delta ),
 \label{hh}
 \end{equation}
 where in general the both angles depend on $(x,t)$.
 At zero applied supercurrent the equilibrium shape of the DW is given by $\delta = \pi/2$ and
 \begin{equation} \label{theta}
 \cos\theta = \pm  \tanh [(x-x_0)/d_w],
 \end{equation}
 where $d_w = \sqrt{A_{ex}/K}$ is the DW width. The above ansatz corresponds to the head-to-head DW, lying in the $xz$-plane.
 The tail-to-tail DW can be obtained by $\theta \to \theta + \pi$.

 Let us consider the behavior of the head-to head DW under the applied supercurrent and
 the presence of  SOC given by the superposition of the Rashba-type
 term $2 m_F \mu_B \bm B_x /M = (0, -\beta_R, 0)$ and the Dresselhaus-type term $2 m_F \mu_B \bm B_x /M = (\beta_D, 0, 0)$.

 First, we follow the Walker's procedure \cite{Schryer1974} by assuming that $\delta = \delta (t)$ and the DW is moving
 according to the time-dependent shift $x_0(t) = \int \limits_0^t v(t')dt'$ in the Eq.(\ref{theta}).
 Substituting this ansatz to the LLG equation we obtain that this type of the solution
 exists only in the absence of Rashba SOC $\beta_R=0$.
 We assume that the distortion of the wall is small during the wall motion, that is
 $\delta = \pi/2 +  \delta_1$, where $| \delta_1| << 1$. In this case taking into account that
 $d_w\nabla_x\theta =  \sin \theta $ for the DW we obtain 
 \begin{align}
 & \partial_t \delta_1 = - \frac{\alpha v}{d_w} - 2 \tilde J^z \beta_D
 \label{LLG_1_lin}
 \\
  &(1+\alpha^2) v  -  u  = \frac{\gamma d_w K_\perp \delta_1}{M} -  2d_w \tilde J^z \alpha \beta_D.
   \label{LLG_2_lin}
 \end{align}
  In this case Eqs.~(\ref{LLG_1_lin}) and (\ref{LLG_2_lin}) yield the  following equation for $v(t)$:
 \begin{equation}
 \partial_t v +  \frac{\gamma\alpha K_\perp}{M(1+\alpha^2)} v = - \frac{2 d_w\gamma K_\perp  \tilde J^z \beta_D}{M(1+\alpha^2)} .
 \label{newton}
 \end{equation}
 Taking into account the initial condition determined by the Eq.~(\ref{LLG_2_lin})
 $(1+\alpha^2)v(t=0)= u - 2 d_w\alpha \beta_D \tilde J^z$, which follows from $\delta_1(t=0)=0$
 we determine the solution of Eq.~(\ref{newton}) in the form:
 \begin{align}
 & v(t) = \Bigl[ u + \frac{2 d_w\tilde J^z \beta_D}{\alpha} \Bigr]
 \frac{e^{-t/t_d}}{(1+\alpha^2)} - \frac{2 d_w \tilde J^z \beta_D}{\alpha},
 \label{v_sol}
  \\
 & \delta(t) = \frac{\pi}{2}+\frac{t_d \Bigl(1-e^{-t/t_d}\Bigr)}{1+\alpha^2}
 \Bigl[ \frac{u \alpha}{d_w}-2 \tilde J^z \beta_D \Bigr].
 \label{delta_sol}
 \end{align}
 where { $t_d=(1+\alpha^2)M/(\alpha \gamma K_\perp)$} is the characteristic time scale.
 The solution for the moving DW expressed by Eqs.~(\ref{v_sol},\ref{delta_sol}) exactly coincides with the solution found for the DW motion in normal ferromagnets under the influence of the adiabatic and nonadiabatic torques \cite{Li2006}. But, nevertheless, there is an important physical difference between the spin-orbit torque, considered here, and the nonadiabatic spin torque. As it can be seen from Eq.~(\ref{torque_so}), the SO torque is equivalent to the torque, generated by an external applied field
 $\gamma \bm H = -4 \mu_B m_F \tilde J_j^z \bm B_j/M $. Consequently, it moves DWs of opposite types (+/- and -/+) to opposite directions as opposed to the action of the nonadiabatic torque, which moves all the DWs in one and the same direction.
 At the same time, it is seen from Eqs.~(\ref{LLG_1_lin}) and (\ref{LLG_2_lin}) that the Rashba SO torque is equivalent to the field perpendicular to the wall plane, therefore it does not move the DW and only distorts it.
 The solution (\ref{v_sol},\ref{delta_sol}) is only valid for small enough electric and, correspondingly, spin currents, applied to the system. If the current is large enough, the condition $ |\delta_1| \ll 1$ is violated and Eqs.~(\ref{LLG_1_lin}) and (\ref{LLG_2_lin}) are not valid. It was shown \cite{Tatara2004} that in this regime for $\tilde J^z > \tilde J_{crit}$ the DW can be moved even by the adiabatic torque only.

\begin{figure}[!tbh]
 \centerline{\includegraphics[clip=true,width=2.8in]{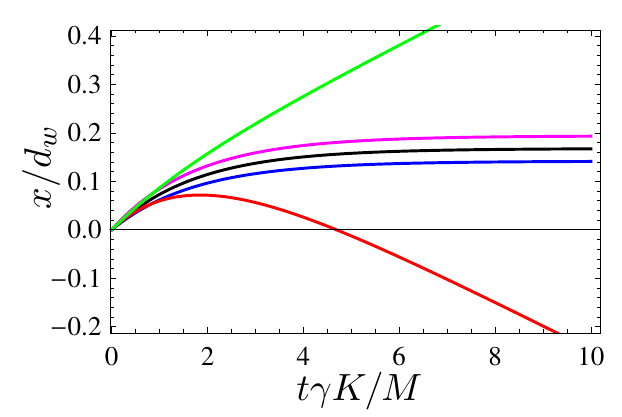}}
 \caption{The displacement of the DW as a function of time in the regime below the threshold current $\tilde J_{crit}$.
 $\tilde \beta_R = \tilde \beta_D = 0$ (black), $\tilde \beta_R = -0.05, \tilde \beta_D = 0$ (blue), $\tilde \beta_R = 0.05, \tilde \beta_D = 0$ (pink),
 $\tilde \beta_R = 0, \tilde \beta_D = -0.05$ (green), $\tilde \beta_R = 0, \tilde \beta_D = 0.05$ (red). The dimensionless parameter $\tilde \beta_{R,D} = \beta_{R,D}d_w M/\mu_B$.
 The other parameters are $K_\perp/K=3.0$, $\alpha = 0.2$ and $\tilde J^z = -0.1 K d_w$ for all the curves.}
 \label{small}
 \end{figure}

 We consider the regime of arbitrary values of the applied current numerically by solving Eq.~(\ref{Eq:Dspin}) together with the expressions for the torque Eqs.~(\ref{torque_adiabatic})-(\ref{torque_so}) and the effective field $H_{eff}$, found from Eq.~(\ref{anisotropy}). The results for the case of small enough applied currents, when our analytical solutions are valid, are represented in Fig.~\ref{small}. The figure demonstrates the displacement of the DW center as a function of time. The black curve corresponds to the case of no spin-orbit torque. The blue and pink curves are for the Rashba case $\beta_R \neq 0 , \beta_D = 0$.
 They demonstrate that the Rashba spin-orbit torque does not move the DW in this case, as it was mentioned above. The green and red curves demonstrate the influence of the Dresselhaus SO torque on the DW motion. In agreement with our analytical calculations, the numerics gives that at $t \gg t_d$ the DW moves with the constant velocity. The direction of the motion is determined by the sign of $\beta_D$ or, in other words, by the sign of the effective magnetic field. In this case it is possible that the DW reverses the direction of its motion if the adiabatic spin torque tends to displace it in the direction opposite to the one dictated by the effective field.
 This case is illustrated by the red curve in Fig.(\ref{small}).

\begin{figure}[!tbh]
   \centerline{\includegraphics[clip=true,width=2.8in]{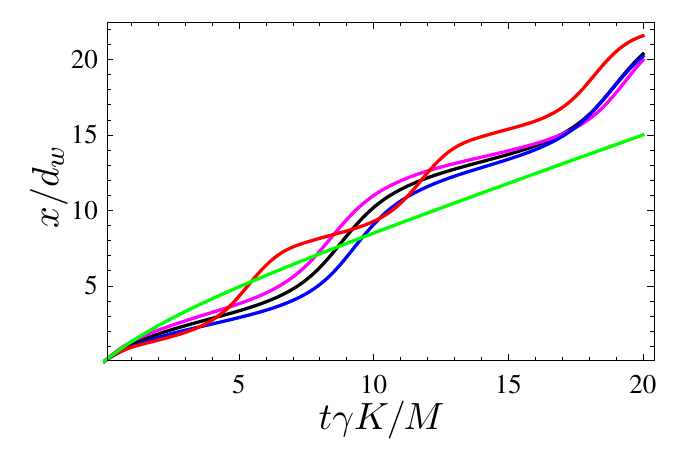}}
   \caption{The displacement of the DW as a function of time in the regime above the threshold current
   $\tilde J^z > \tilde J_{crit}$.
   $\tilde \beta_R = \tilde \beta_D = 0$ (black), $\tilde \beta_R = -0.05, \tilde \beta_D = 0$ (blue), $\tilde \beta_R = 0.05, \tilde \beta_D = 0$ (pink),
   $\tilde \beta_R = 0, \tilde \beta_D = -0.05$ (green), $\tilde \beta_R = 0, \tilde \beta_D = 0.05$ (red).
   The other parameters are $K_\perp/K=3.0$, $\alpha = 0.2$ and $\tilde J^z = -1.5 K d_w$ for all the curves.}
   \label{large}
  \end{figure}

The regime of large applied currents $\tilde J^z > \tilde J_{crit}$, when the DW can be moved by the adiabatic torque only,
is shown in Fig.~\ref{large}. We have obtained that the value of $\tilde J_{crit}$ is rather close to $ K d_w$.
Therefore the critical electric current density is of the order $eK d_w/\hbar \sim 10^{10}$ A/m$^2 $
which is an order of magnitude larger the Josephson critical current obtained in experiment\cite{Singh2016}.
Again, the black curve in Fig.~\ref{large} shows the displacement of the DW in the absence of the SO torques.
The initial dynamics of the DW at small $t$ coincides with Fig.~\ref{small}, but at larger values of $t$ the situation changes, so that in
this regime the DW moves, but its velocity is not constant.
 The Rashba SO torque does not cause any essential influence on the DW dynamics, as in the case of the small applied currents.
But the effect of Dresselhaus SOC is significant and at the first glance unexpected.
 Indeed, as shown in  Fig.~\ref{small} the torque generated by this type of SOC
 e.g. for $\beta_D <0$ moves the DW to the direction $x>0$.
 But in the above-threshold regime it can also  reduce the averaged DW velocity (green curve in Fig.~\ref{large}), that is the combined action of the adiabatic ST torque and SO torque cannot be viewed just as a simple sum of independent motions due to the both reasons.
 Vice versa, the SO torque generated at $\beta_D >0$, which by itself tends to move the DW to the direction $x<0$, can slightly enhance the average DW velocity, as it is demonstrated by the red curve.

\begin{figure}[!tbh]
 \begin{minipage}[b]{\linewidth}
   \centerline{\includegraphics[clip=true,width=2.6in]{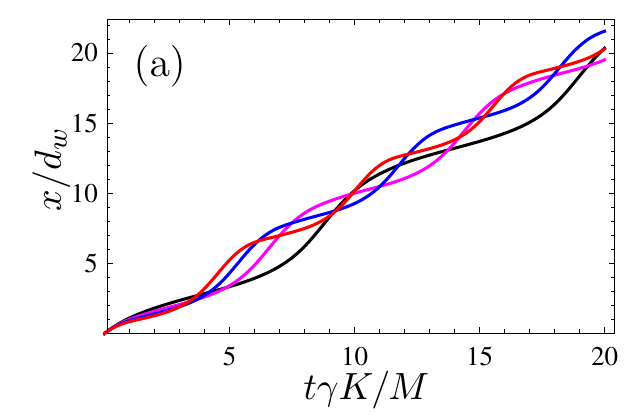}}
   \end{minipage}\hfill
   \begin{minipage}[b]{\linewidth}
   \centerline{\includegraphics[clip=true,width=2.5in]{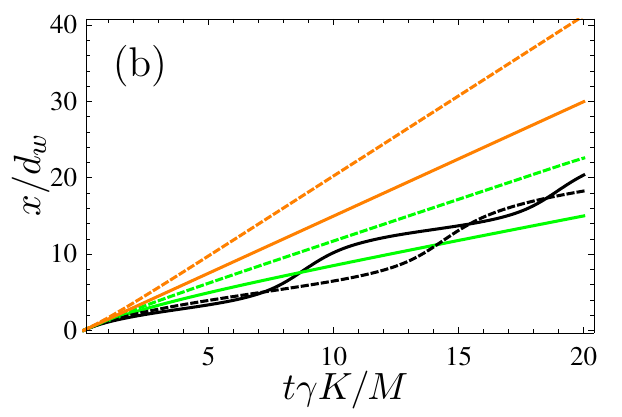}}
   \end{minipage}
   \caption{The displacement of the DW as a function of time in the regime above the threshold current
   $\tilde J_{crit}$ for the case of $\tilde \beta_R = 0$ and different values of the Dresselhaus SO coupling.
   (a) $\tilde \beta_D = 0$ (black), $0.025$ (pink), $0.05$ (blue), $0.075$ (red); (b) $\tilde \beta_D = 0$ (black),
   $-0.025$ (dashed black), $-0.05$ (green), $-0.075$ (dashed green), $-0.1$ (tan), $-0.15$ (dashed tan).
   The other parameters are as in Fig.~\ref{large}.}
 \label{large2}
 \end{figure}

The influence of the Dresselhaus SOC on the DW average velocity is represented in Fig.~\ref{large2} in more detail. Fig.~\ref{large2}(a) demonstrates the displacement of the DW as a function of time $t$ for several values of $\beta_D >0 $.
It is seen that there is a weak increase of the average velocity at $\beta_D>0$, but the more important and pronounced
 effect is that increasing $\beta_D$ leads to the decrease of the velocity oscillation period.
 The case $\beta_D<0$ is shown in Fig.~\ref{large2}(b), where once can see that
  the dependence of the average DW velocity on $\beta_D$ is nonmonotonous.
  While at smaller values of $|\beta_D|$ the average velocity is indeed reduced with respect to the case $\beta_D = 0$,
   at larger  values of $|\beta_D|$ the velocity starts to increase and exceed its value at $\beta_D=0$ considerably.

This behavior can be understood in the framework of the analogy between the SO torque and the magnetic-field induced torque. For the situation when the DW moves under the combined action of the current-induced torque and field-induced torque it is known that the steady motion of the DW with $\dot \delta = 0$ is only possible for a range of fields and currents \cite{Mougin2007}. The lines in the
{ $(\tilde J^z,H)$ } -plane, separating the regions of steady motion and precession motion $\dot \delta \neq 0$, are called by the Walker-like stability lines \cite{Mougin2007}. This limit condition for the steady motion is strictly equivalent to the Walker breakdown \cite{Schryer1974} condition in the case where only an external magnetic field is applied. For the problem under consideration the increase of $\beta_D$ absolute value at fixed current is equivalent to the increase of the applied field (at fixed current). When at zero $\beta_D$ the system is in the precession regime, as in Fig.~\ref{large}, the increase of $|\beta_D|$ at $\beta_D<0$ moves the system towards the steady motion region, where the wall velocity is higher. Therefore, the transition from the precession regime to the steady regime in Figs.\ref{small} and \ref{large} is analogous to crossing the Walker-like stability lines for problem of DWs motion under the combined action of the current-induced torque and field-induced torque.

  Let us now consider the Neel DW. In this case the combination of the anisotropy energy and the exchange energy takes the form:
 \begin{equation}
 F= \frac{1}{2}\int d^3 r \Biggl[ K_\perp m_z^2  - K m_y^2 + A_{ex} ( \nabla_x \bm m )^2  \Biggr].
 \label{anisotropy_Neel}
 \end{equation}
 It is convenient to parametrize the magnetization as:
 \begin{eqnarray}
 \bm m = (\sin \theta \sin \delta, \cos \theta, \sin \theta \cos \delta).
 \label{neel}
 \end{eqnarray}
 At zero applied supercurrent the equilibrium magnetization profile is described by
 Eqs.~(\ref{theta}) and $\delta=\pi/2$. It can be shown that the problem of the Neel DW motion in the presence of the Rashba SO coupling is mathematically equivalent to the considered above motion of the head-to-head DW in the presence of the Dresselhaus SO coupling with the substitution $\beta_D \to -\beta_R$. Therefore, in this case the Rashba SO torque plays the part of the field-induced torque moving DWs.

 The above analysis demonstrates that the dynamics of a DW under an applied supercurrent depends strongly
 (i) on the particular type of the DW and (ii) on the particular type of the SO coupling, which induces the spin-orbit torque.
 The stationary motion of the DWs induced by small supercurrents is possible even in the absence of the nonadiabatic torque
 if the spin-orbit torque is present in the system.

 Due to the presence of the Gilbert damping the motion of a DW by a supercurrent is not a disspationless process.
 Interestingly the DW motion generates voltage  across the junction in the regime when the charge current is fixed
   but its magnitude is smaller than the Josephson critical current of the system. In this situation the voltage can manifest itself as an additional step at the current-voltage characteristics of the junction at $j<j_c$, where $j_c$ is the critical current of the junction.
  The voltage amplitude $V$ can be roughly estimated from the balance of the energy dissipation rate in the magnetic subsystem due to the Gilbert damping and the power put in by the current source. The characteristic energy dissipation rate can be estimated as $\dot F \sim \Delta F/t_d$, where $\Delta F$ is the difference between the free energies of the equilibrium state of the DW at zero current and the nonequilibrium state
  of the distorted wall in the presence of the current. Our quasiequilibrium consideration of the DW dynamics is strictly valid only if $eV$ is small with respect to the characteristic inverse time scale of the problem $\gamma K/M$. For small distortions of the DW $\Delta F$ can be obtained as follows:
 \begin{align}
 \Delta F = \frac{1}{2}\int d^3 r
 \left[ K_\perp \sin^2 \theta + A_{ex}(\nabla_x\theta)^2 \cos^2 \theta \right]
 \delta_1^2
 \label{GDrate}
 \end{align}
 Substituting the equilibrium profile of the DW $\theta(x)$ given by the Eq.(\ref{theta}) into Eq.~(\ref{GDrate}), we obtain:
 \begin{eqnarray}
 \Delta F = S_p d_w (K_\perp+K/3) \delta_1^2,
 \label{GDrate_1}
 \end{eqnarray}
 where $S_p$ is the cross-section area of the ferromagnet. The voltage, generated at the Josephson junction can be estimated as
 $V \sim \dot F/S_p j_c$ which yields
 \begin{equation}
 V \sim \frac{\gamma \delta_1^2 \alpha d_w K_\perp (K_\perp+K/3)}{j_c M \hbar},
 \label{voltage}
 \end{equation}
 where we have assumed that $\alpha \ll 1$.

  For estimations we use the material parameters of the CrO$_2$ nanostructures \cite{Zou2007} which are the promising systems for the
 dissipationless spintronics \cite{Singh2016}. Taking the maximal Josephson current density through the CrO$_2$ nanowire to be
 $j_c\sim 10^9\; A/m^{2}$, the saturation magnetization $M = 4.75\times 10^5\; A/m$, $d_w = 10^{-6} cm$,
 $K = 1.43 \times 10^5 erg/cm^3$ and $K_\perp = 3 K$, we obtain $V \sim 0.1 \delta_1^2 [mV] $ , where we took into account
 the typical values of the Gilbert damping $\alpha \sim 0.01$.
 The amplitude of DW distortion angle can be varied in wide limits, e.g. $\delta_1^2 \sim 10^{-4}-10^{-3}$ for
 the red curve in Fig.\ref{small}, $\delta_1^2 \sim 10^{-3}$ for the green curve in Fig.\ref{small}
 and $\delta_1^2 \sim 10^{-1}$ for the dashed green curve in Fig.\ref{large2}b. The estimated values of the induced voltage $V$ are small with respect to the characteristic superconducting scales $\sim 0.1 [mV]$ for Al superconductors, therefore our assumption of quasiequilibrium quasiparticle distribution works rather well. From the other hand, the strict calculation of the voltage induced at the Josephson junction requires accounting for dynamics of the superconducting phase induced by the DW motion in the current-phase relation. This is beyond the scope of the present paper and will be done elsewhere.

 \section{Conclusion}
 \label{Sec:Conclusions}
 To conclude, we have calculated the spin transfer torques acting on the magnetic textures
  from the spin-polarized superconducting current flowing through the ferromagnetic material. For this we take the advantage
  of the widely used adiabatic approximation, bringing it from the realm of single-electron dynamics into the
   field of superconductivity governed by the propagation of the spin-triplet Cooper pairs generated at the SC/FM interface.
  This approximation enables us to find the analytical expression for the spin torques in the most general case of the spin texture and
  develop the efficient formalism of the generalized quasiclassical theory for calculating
   the charge and spin supercurrents through the   inhomogeneous magnetic systems.
    We show that the supercurrent-driven dynamics of DWs crucially depends on the type and magnitude of the
   spin-orbital coupling. The obtained results demonstrate that the DW motion by the supercurrent is a phenomenon
    realistic for the recently developed Josephson junctions through CrO2 nanowires.

 \section{Acknowledgements}
 This work was supported by the Academy of Finland  Research Fellow (Project
No. 297439)and RFBR Grant No.18-02-00318.
We thank Jan Aarts and Tero Heikkila for interesting discussions that initiated this project.

	\appendix
	\section{ Calculation of the effective field Eq.(\ref{Eq:EffectiveField}) }
	\label{Apx:1}
	
     From the Eqs.(\ref{josephson:Spin},\ref{Eq:JosephsonEnergy}) we obtain  	
     \begin{equation}
     \frac{\partial F_J}{\partial \bm M} = \frac{2 \tilde J_{x}^z}{M} \frac{\delta \int_{-d/2}^{d/2}Z_x dx }{\delta \bm m}
     \end{equation}     	
		
	  Let's consider the following form of the unitary matrix $\hat U = \exp\{-i\sigma_x (\frac{\delta}{2}+\frac{\pi}{4})\}\exp\{-i\sigma_y (\frac{\theta}{2}+\frac{\pi}{4})\}$,
	 which yields the texture part of the gauge field $\bm Z^m = -\cos\theta\nabla\delta/2$, where
	 $m_x = \cos\theta$ and $\tan\delta = m_z/m_y$ so that $\nabla\delta = (m_y\nabla m_z - m_z\nabla m_y)/m_\perp^2$,
	 where $m_\perp =\sqrt{ m_y^2 + m_z^2}$.
	 Then we get
	 \begin{align}
	 & \frac{\delta} {\delta \bm m_x} \int_{-d/2}^{d/2} \bm Z^{m} dx =
     -\frac{\bm x}{2m_\perp^2} ( m_y\nabla m_z - m_z\nabla m_y )
	  \\
	  & \frac{\delta} {\delta \bm m_y} \int_{-d/2}^{d/2} \bm Z^{m} dx =
	  -\frac{\bm y}{2m_\perp^2}  m_z\nabla m_x
	  \\
	  & \frac{\delta} {\delta \bm m_z} \int_{-d/2}^{d/2} \bm Z^{m} dx = \frac{\bm z}{2m_\perp^2}  m_y\nabla m_x
	 \end{align}
	
     Hence
     \begin{align}
     & 2 \left( \bm M\times \frac{\delta \int_{-d/2}^{d/2}Z_x dx }{ M \delta \bm m}\right) _y =\frac{1}{m_\perp^2} \times
         \\ \nonumber
	  &  [ m_z (m_z\nabla_x m_y -m_y\nabla_x m_z ) - m_y m_x\nabla_x m_x ] =
	  \nabla_x m_y .
     \end{align}
     Treating analogously  other components and the spin-orbital part of the gauge field we get
         \begin{align}
     & 2 \left( \bm M\times \frac{\delta \int_{-d/2}^{d/2}Z^{m}_x dx }{ M \delta \bm m}\right)  =  \nabla_x \bm m
     \\
     & 2 \left( \bm M\times \frac{\delta \int_{-d/2}^{d/2}Z^{so}_x dx }{M \delta \bm m}\right) = -2m_F \bm B_x \times \bm M
     \end{align}       	
    Combining that into the total effective field yields Eq.(\ref{Eq:EffectiveField}).

%

 \end{document}